\documentclass[twocolumn,showpacs,,prl]{revtex4}
\usepackage{psfrag}
\usepackage{graphicx}
\usepackage{amssymb}
\usepackage{amsmath,amsfonts,latexsym}

\newcommand{\be}{\begin{equation}}
\newcommand{\ee}{\end{equation}}
\newcommand{\bea}{\begin{eqnarray}}
\newcommand{\eea}{\end{eqnarray}}

\newcommand{\p}{\partial}

\newcommand{\lp}{\left(}
\newcommand{\rp}{\right)}
\renewcommand{\phi}{\varphi}

\renewcommand{\vec}[1]{{\mathbf #1}}
  
\begin{document}

\title{Pattern Formation as a Signature of Quantum Degeneracy in a Cold 
Exciton System}
\author{L.\,S.~Levitov$^1$,  B.\,D.~Simons$^2$, and L.\,V.~Butov$^3$}
\address{$^1$Department of Physics,
Center for Materials Sciences \& Engineering,
Massachusetts Institute of Technology, 77 Massachusetts Ave,
Cambridge, MA 02139}
\address{$^2$Cavendish Laboratory, Madingley Road, Cambridge CB3 OHE, UK}
\address{$^3$Department of Physics, University of California San Diego, 
La Jolla, CA 92093-0319}

\begin{abstract}
The development of a Turing instability to a spatially modulated state
in a photoexcited electron-hole system is proposed as a novel signature of 
exciton Bose statistics.
We show that such an instability, which is 
driven by kinetics of exciton formation, can result from stimulated 
processes that build up near quantum degeneracy. In the spatially uniform
2d electron-hole system, the instability leads to a triangular lattice 
pattern while, at an electron-hole interface, a periodic 1d pattern 
develops. We analyze the mechanism of wavelength selection,
and show that the transition
is abrupt (type I) for the uniform 2d system, and continuous (type II)
for the electron-hole interface.
\end{abstract}


\maketitle

Recently, striking spatial photoluminescence (PL) patterns~\cite{Butov02,%
Snoke02,Butov03,Snoke03,Rapaport}, which span macroscopic scales in 
excess of $100\,\mu$m, have been observed in photoexcited AlGaAs/GaAs 
quantum well (QW) structures. In addition to concentric rings and `bright 
spots', the electron-hole system exhibits an abrupt transition at 
ca.~$2\,$K in which the outermost ring `fragments' into regularly 
spaced beads of high PL intensity~\cite{Butov02,Butov03}. While the gross
features of PL have been explained within classical framework,
attributing the internal rings to nonradiative exciton transport
and cooling~\cite{Butov02} and the outermost rings and 'bright spots'
to macroscopic charge separation~\cite{Butov03,Snoke03,Rapaport},
the origin of the instability remains unidentified.

Spatially modulated exciton density 
is not to be expected in QW
system designed so that excitons interact repulsively
as electric dipoles~\cite{Butov02} 
and thus do not form droplets~\cite{Keldysh}. 
The macroscopic character of ring fragments,
of  $10-30\,\mu{\rm m}$ in size
and containing about $10^4$ excitons each, 
and the abrupt temperature dependence,
call for an explanation involving a symmetry-breaking 
instability of a homogeneous state to a patterned state. Such behavior is
reminiscent of the instability predicted by Alan Turing~\cite{Turing52} to 
occur in a reaction-diffusion system.
The Turing instability is known to occur in certain chemical 
reactions~\cite{Castets90,Kepper91,Ouyang91}, and is also believed to be 
relevant for pattern formation in biological systems~\cite{Murray89}.

In this work we propose a mechanism, based on the kinetics of exciton 
formation from optically excited electrons and holes, that can lead to 
instability in the exciton system. Interestingly, these kinetic 
effects 
become especially strong 
in the regime near \emph{exciton quantum 
degeneracy}, due to stimulated enhancement of the electron-hole binding
rate. 
The transition to a state with a spatially modulated 
exciton density reveals itself in the spatial PL pattern, and presents a
directly observable signature of degeneracy. 

Although, in itself, an 
observation of an instability does not constitute unambiguous evidence
for degeneracy, it may complement other manifestations 
discussed in the 
literature, such as changes in exciton 
recombination~\cite{Butov99} and scattering~\cite{Butov01} rates,
in the PL spectrum~\cite{Timofeev}, 
absorption~\cite{Johnsen01} and PL angular distribution~\cite{Keeling03}.
%
%
While linking the observed instability with degeneracy is 
premature, our main aim here is to present the Turing 
instability from 
a broader viewpoint,
as a novel and quite general 
effect of quantum kinetics that can help to identify the regime of 
Bose-Einstein condensation (BEC).


Here we consider a transport theory~\cite{Butov03} formulated in 
terms of electron, hole, and exciton densities $n_{\rm e,h,x}$ obeying a 
system of coupled nonlinear diffusion equations:
%
\begin{eqnarray}
&& {\rm (e)}\quad \p_t n_{\rm e}=D_{\rm e}\nabla^2 n_{\rm e}-w n_{\rm e}
n_{\rm h}+J_{\rm e}\nonumber\\\label{eq:Diff}
&& {\rm (h)}\quad \p_t n_{\rm h}=D_{\rm h}\nabla^2 n_{\rm h}-w n_{\rm e}
n_{\rm h}+J_{\rm h}\\\nonumber
&& {\rm (x)}\quad \p_t n_{\rm x}=D_{\rm x}\nabla^2 n_{\rm x}+w n_{\rm e} 
n_{\rm h}-\gamma n_{\rm x}.
\end{eqnarray}
The nonlinear couplings account for exciton formation from free 
electron and hole 
binding.
Here, no attempt has been made to describe the
detailed and complicated density and temperature dependence of the physical 
parameters entering the model, nor to account for 
nonequilibrium
exciton energy distribution and
cooling due to phonon emission~\cite{Ivanov,Butov01}.
Instead, we adopt a more phenomenological approach and assume that, 
as a result of cooling, the system can be described by an effective 
temperature, which leaves the densities $n_{\rm e,h,x}$ as the only important 
hydrodynamical variables. The sources $J_{\rm e}$ and $J_{\rm h}$ in 
Eq.~(\ref{eq:Diff}) describe the carrier photo-production,
as well as the leakage current in the QW 
structure. 

In general, one can expect the electron-hole binding rate $w$ 
and, to a lesser extent, the exciton recombination rate $\gamma$ 
to depend sensitively on the local exciton density $n_{\rm x}$. 
Of the several mechanisms that could lead to such a dependence 
at low temperatures close to exciton degeneracy, perhaps the most 
important in the present context involve stimulated electron-hole 
binding processes mediated by phonons. These 
processes enhance the binding rate by a factor $f=1+N_E^{\rm eq}$, where 
$N_E^{\rm eq}$ denotes the occupation of exciton states. 
In thermal equilibrium, and at low temperatures, 
one can ignore the reverse processes of exciton dissociation: 
The binding energy, carried away by phonons, is much larger
than $k_{\rm B}T$.

For a degenerate exciton gas with $N_{E=0}>1$, the dominant process 
involves scattering into the ground state and the stimulated enhancement 
factor is expressed as
\begin{equation}\label{eq:f-factor}
f=e^u \,,\quad 
u\equiv \frac{n_{\rm x}}{n_0(T)}\,,\quad n_0(T)=\frac{2 g m_{\rm x}
k_{\rm B}T}{\pi\hbar^2}.
\end{equation}
Here $m_{\rm x}\simeq 0.21m_0$ represents the exciton mass, and $g$ denotes 
the degeneracy (in the indirect exciton system, the exchange interaction is 
extremely weak, and $g=4$). Equivalently, when reparameterized through its 
dependence on temperature, $u\equiv T_0/T$ where $T_0=\left(\pi\hbar^2/2
gm_{\rm x}k_{\rm B}\right) n_{\rm x}$ is the degeneracy temperature. At  
$T\sim T_0$ (equivalently $n_{\rm x}\sim n_0$), there is a crossover from 
classical to quantum Bose-Einstein statistics, and the stimulated enhancement 
factor $f$ increases sharply. 

Qualitatively, the stimulated transition mechanism for hydrodynamic
instability can be understood as follows:
A local fluctuation in the exciton density leads to an increase in the 
stimulated electron-hole binding rate. The associated depletion of the 
local carrier concentration causes neighboring carriers to stream 
towards the point of fluctuation presenting a mechanism of positive
feedback.
The wavelength, determined by the most unstable harmonic of the density,
characterizes the lengthscale of spatial modulation in the nonuniform state.

Before turning to the analysis of instability, it is useful to discuss 
intrinsic constraints on the dynamics (\ref{eq:Diff}) due to electric 
charge and particle number conservation. These are obtained by considering 
the linear combinations ${\rm (e)-(h)}$, ${\rm (e)+(h)+2(x)}$ of the 
transport 
equations (\ref{eq:Diff}). In both cases, the nonlinear term drops out and 
one obtains linear equations
\begin{eqnarray}\label{eq:e-h}
&& \hat L_{\rm e} n_{\rm e} - \hat L_{\rm h} n_{\rm h}
=J_{\rm e}-J_{\rm h}
\\ \label{eq:e+h+2x}
&&
\hat L_{\rm e} n_{\rm e} 
+ \hat L_{\rm h} n_{\rm h}
+ 2\hat L_{\rm x} n_{\rm x} +2\gamma n_{\rm x} = J_{\rm e}+J_{\rm h}
\end{eqnarray}
with 
$\hat L_{\rm e(h,x)}=\p_t-D_{\rm e(h,x)}\nabla^2$. Note that, since the 
origin of the relations (\ref{eq:e-h},\ref{eq:e+h+2x}) 
is routed in conservation laws, they are robust and 
insensitive to the exact form of the electron-hole binding term.

\begin{figure}[hbt]
\psfrag{(k/l)**2}{$({\bf k}/k_\ast)^2$}
\psfrag{u>uc}{$u>u_c$}
\psfrag{u<uc}{$u<u_c$}
\psfrag{u=uc}{$u=u_c$}
\psfrag{k1}{${\bf k}^{(1)}$}
\psfrag{k2}{${\bf k}^{(2)}$}
\psfrag{k3}{${\bf k}^{(3)}$}
\centerline{
\includegraphics[width=0.75\linewidth,angle=0]{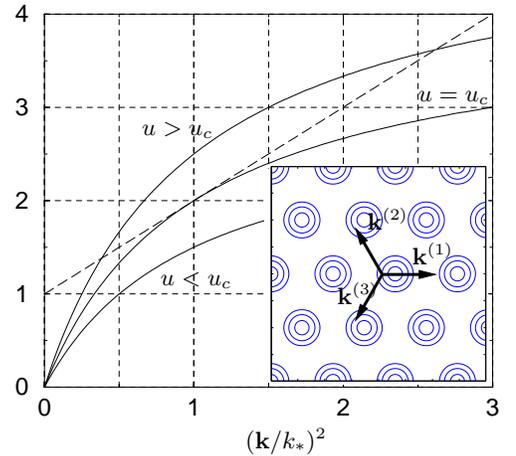}}
\vspace{-0.1in}
\caption{Graphical solution of Eq.~(\ref{eq:f(k)=0}) that selects the most 
unstable wavelength. {\it Inset:} The $3$-fold symmetric star of wavevectors 
describing the modulation near the instability threshold and the 
corresponding triangular pattern of exciton density variation.}
\label{fig:2d}
\end{figure}

Initially, let us consider a system in which the sources are constant 
$J_{\rm e}({\bf r})=J_{\rm h}({\bf r})\equiv J$, realized by a spatially 
extended photoexcitation.
In this case, ignoring the dependence of the 
recombination rate $\gamma$ on 
density, we have 
\begin{equation}\label{eq:uniformstate}
\bar n_{\rm x}=J/\gamma \,,\quad \bar n_{\rm e,h}=\lp J/w(\bar n_{\rm x})
\rp^{1/2}.
\end{equation}
The stability of the system
can be assessed by linearizing Eqs.~(\ref{eq:Diff}) about the uniform 
solution (\ref{eq:uniformstate}) with a harmonic modulation 
$\delta n_{\rm e,h,x} \propto e^{\lambda t}e^{i\vec k\cdot\vec r}$. 
Using 
(\ref{eq:e-h},\ref{eq:e+h+2x}) and writing $\hat L_{\rm e}\delta 
n_{\rm e}=\hat L_{\rm h}\delta n_{\rm h}=- \hat L_{\rm x}\delta n_{\rm x}
-\gamma \delta n_{\rm x}$,
one can express $\delta n_{\rm h,x}$ in terms of $\delta n_{\rm e}$ and obtain
\begin{equation}
L_{\rm e}(\lambda,\vec k)+\gamma\frac{\bar{n}_{\rm x}}{\bar{n}_{\rm e}}
\lp 1+\frac{L_{\rm e}(\lambda,\vec k)}{L_{\rm h}(\lambda,\vec k)}\rp =
\frac{\gamma u L_{\rm e}(\lambda,\vec k)}{L_{\rm x}(\lambda,
\vec k)+\gamma}
\label{eq:2Dlinearized}
\end{equation}
where 
$u=d\ln w/d\ln \bar{n}_{\rm x}$ is
evaluated 
at the steady state (\ref{eq:uniformstate}), and $L_{\rm e(h,x)}(\lambda,
\vec k)=\lambda+D_{\rm e(h,x)}\vec k^2$. Solving Eq.~(\ref{eq:2Dlinearized}),
one obtains the growth rate dispersion $\lambda(\vec k)$ from which one
can infer from the stability criterion, ${\rm Re}\,\lambda <0$, that the
system becomes unstable when
%
\begin{equation}\label{eq:f(k)=0}
({\bf k}\ell_{\rm x})^2+r=u\frac{({\bf k}\ell_{\rm x})^2}{1+
({\bf k}\ell_{\rm x})^2},
\end{equation}
where $\ell_{\rm x}=\sqrt{D_{\rm x}/\gamma}$ denotes the exciton diffusion
length and $r=D_{\rm x}(D^{-1}_{\rm e}+D^{-1}_{\rm h})\bar{n}_{\rm x}/
\bar{n}_{\rm e}$. Eq.~(\ref{eq:f(k)=0}) has solutions if 
%
\begin{equation}\label{eq:dlnw/dlnn}
u\equiv \frac{d\ln w}{d\ln \bar{n}_{\rm x}} \ge u_c=\lp 1+r^{1/2}\rp^2
\end{equation}
as illustrated in Fig.~\ref{fig:2d}. At $u=u_c$, we obtain the most unstable 
wavenumber $k_\ast = r^{1/4}/\ell_{\rm x}$
selected by competition of the stimulated binding and diffusion 
processes. 

The binding rate $w(n_x)$ builds up near degeneracy due to the 
growth of stimulated processes, leading to instability
at temperatures approaching $T_{\rm BEC}$.
(For a weakly interacting Bose gas the transport coefficients, and thus 
the constant $r$, are 
practically insensitive to the degree of exciton degeneracy 
at $T>T_{\rm BEC}$.) 

To what extent are these results insensitive to the origin of the nonlinearity
in the binding rate? If enhanced by intraband Auger processes, 
which transfer the binding energy released in exciton formation 
to other excitons, one expects the 
binding rate $w$ to scale linearly with local exciton density, viz. 
$w(n_{\rm x})=w_0(1+n_{\rm x}/\tilde n_0)$, where $\tilde n_0$ denotes some constant 
involving a ratio of the two-body and three-body cross-section of the electron 
and hole in the presence of excitons. Crucially, in this case, the left hand 
side of Eq.~(\ref{eq:dlnw/dlnn}) is bounded by unity, while the right 
hand side is in excess. Therefore, at least over the parameter 
range considered here, one can infer that a simple linear scaling of the 
binding rate with density 
does not lead to 
instability. Indeed, the instability may be used 
to discriminate against certain mechanisms in the kinetics of exciton 
formation.

Turning to the discussion of the spatial pattern resulting from the 
instability, we note that the wavevector selection determines its 
modulus, but not direction. At threshold $u=u_c$ all modes 
with $|\vec k|=k_\ast$ become unstable simultaneously. The resulting $2$d 
density distribution can be found by considering the effect of mixing  
different harmonics due to higher order terms in  
(\ref{eq:Diff}) expanded in $\delta n_{\rm e,x}$ about the uniform state. 
Since these equations contain quadratic terms, the favored combination of 
harmonics is a $3$-fold symmetric star $\vec k^{(j)}=
k_\ast (\cos(\textstyle{\frac{2\pi}3} j+\theta),
\sin(\textstyle{\frac{2\pi}3} j+\theta)),\ j=1,2,3$,
%
%
with the parameter $\theta$ describing the degeneracy with respect to 
$2$d rotations. This leads to a density distribution $\delta n\propto
\sum_je^{\pm \vec k_j\cdot\vec r}$ with maxima arranged in a triangular 
lattice. 

On symmetry grounds, since the triangular lattice pattern
is stabilized by quadratic terms, the mean field analysis predicts that
the transition to the modulated state in this case is abrupt, of a type I kind.
Indeed, the triangular lattice geometry arises in various 
$2$d pattern selection problems, from B\'enard convection cells~\cite{Benard} 
to the mixed state of type II superconductors~\cite{typeII}.

To explore the application of these ideas to the $1$d modulation seen in 
exciton rings~\cite{Butov02,Butov03}, 
one must first determine the profile of the 
uniform distribution. The rings represent an interface 
between regions populated by electrons and holes at which they bind 
to form excitons. The steady state is maintained by a constant flux of 
carriers arriving at the interface. The parameter regime which 
is both relevant and simple to analyze is that of long exciton 
lifetime $\gamma^{-1}$ where the diffusion length $\ell_{\rm x}$
exceeds the range of the 
electron and hole profile overlap. In this case, approximating the source of 
excitons by a straight line
$c\delta(x)$, where $c$ is the total carrier flux and $x$ 
is the coordinate normal to the interface, the exciton density profile is 
given by $(c\ell_{\rm x}/2D_{\rm x})e^{-|x|/\ell_{\rm x}}$.
Accordingly, one can seek the electron and hole profile treating 
$w(n_{\rm x})$ as constant and restoring its dependence on $n_{\rm x}$ 
later when turning to the instability. The profiles can be inferred from 
two coupled nonlinear diffusion equations 
\begin{eqnarray}
D_{\rm e(h)}\partial_x^2 n_{\rm e(h)} =w n_{\rm e}n_{\rm h},
\label{eq:Diff_eh}
\end{eqnarray}
with the 
boundary condition:
$D_{{}^{\rm e}_{\rm h}}\partial_x n_{{}^{\rm e}_{\rm h}}
|_{\pm\infty}=\pm c\,\theta(\pm x)$.
%
From Eq.~(\ref{eq:e-h}) one obtains $D_{\rm e}n_{\rm e}-D_{\rm h} n_{\rm h}=cx$
which allows the elimination of $n_{\rm h}$. 
Applying the rescaling $n_{\rm e(h)}=c\ell\, g_{\rm e(h)}/D_{\rm e(h)}$, 
where $\ell=(D_{\rm e}D_{\rm h}/wc)^{1/3}$, one obtains
%
\begin{equation}
\partial_{\tilde x}^2\ g_{\rm e}=g_{\rm e}(g_{\rm e}-\tilde x)
,\quad \tilde x\equiv x/\ell.
\label{eq:Diff_ge}
\end{equation}
From the rescaling one can infer that the electron and hole profiles 
overlap in a range of width $\ell\sim c^{-1/3}$ while 
\begin{eqnarray*}
g_{\rm e}(|x|\gg \ell) = \tilde x\,\theta(\tilde x)+{\cal O}\lp 
|\tilde x|^{-1/4}e^{-2|\tilde x|^{3/2}/3}\rp.
\end{eqnarray*}

\begin{figure}[hbt]
\psfrag{x/l}{{$x/\ell$}}
\psfrag{U}{$U_{\rm eff}$}
\psfrag{Y}{$\psi\,(\times 10)$}
\centerline{
\includegraphics[width=0.7\linewidth,angle=0]{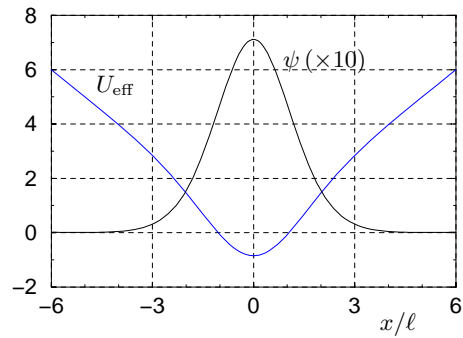}
}
\vspace{-0.1in}
\caption{Numerical solution of the (normalized) bound state wavefunction 
$\psi(x)$ 
representing fluctuation in the electron density, together with the 
effective potential $U_{\rm eff}=\bar{g}_{\rm e}+\bar{g}_{\rm h}-a_0
\,\bar{g}_{\rm e}\bar{g}_{\rm h}$ 
of the Schr\"odinger-like 
equation~(\protect\ref{eq:se}).\vspace{-0.1in}}
\label{fig:wav_pot}
\end{figure}

Although Eqs.~(\ref{eq:Diff_eh}) are nonlinear, their diffusive character 
does not straightforwardly admit a spatial instability: A fluctuation in 
the position of the interface initiates an increased 
electron-hole flux which, in time, restores the uniform distribution. 
However, if one restores the dependence of the binding rate on exciton 
density, the same mechanism of positive feedback which characterized the 
instability in the uniform system becomes active.
To explore the instability, one may again expand linearly in
fluctuations around the spatially uniform solution, $g_{\rm e}(x,y)=
\bar{g}_{\rm e}(x)+\delta g_{\rm e}(x)\,e^{iky}$ (similarly $g_{\rm h}$ 
and $g_{\rm x}$), where $y$ is the coordinate along the interface
and $\bar{g}_{\rm e}(x)$ denotes the uniform profile obtained from 
Eq.~(\ref{eq:Diff_ge}). With $\ell_{\rm x}\gg \ell$, the exciton density 
remains roughly uniform over the electron-hole interface. Denoting this 
value by $\bar{n}_{\rm x}(0)$, in the vicinity of the interface, one may
again develop the linear expansion $w[n_{\rm x}]\simeq w[\bar{n}_{\rm x}(0)]
\,(1+u\delta n_{\rm x}/\bar{n}_{\rm x}(0))$ where, as before, $u=d\ln w/
d\ln n_{\rm x}$. Noting that Eq.~(\ref{eq:e-h}) enforces the steady state 
condition $\delta g_{\rm e}=\delta g_{\rm h}$, a linearisation of 
Eqs.(\ref{eq:Diff})
obtains the Schr\"odinger-like equation
\begin{eqnarray}
\left[-\partial_{\tilde{x}}^2+(\ell k)^2+\bar{g}_{\rm e}+\bar{g}_{\rm h}\right]
\delta g_{\rm e}+\frac{u}{\bar{g}_{\rm x}(0)}\bar{g}_{\rm e}\bar{g}_{\rm h}
\,\delta g_{\rm x}=0,
\label{eq:se}
\end{eqnarray}
together with the condition on the Fourier components,
\begin{eqnarray}
\delta g_{\rm x}(q)=-\delta g_{\rm e}(q)\left(1-\frac{1}
{Q^2+(q\ell_{\rm x})^2}\right), 
\label{eq:xe}
\end{eqnarray}
with $Q^2\equiv (k\ell_{\rm x})^2+1$. 
Now, since the product 
$\bar{g}_{\rm e}\bar{g}_{\rm h}$ is strongly peaked around the interface, 
the typical contribution from the last term in (\ref{eq:se}) arises from 
Fourier elements $q\ell\sim 1$. Then, with $\ell_{\rm x}\gg \ell$, the second 
contribution to $\delta g_{\rm x}(q)$ can be treated as a small perturbation 
on the first and, to leading order, neglected, i.e. $\delta g_{\rm x}\simeq 
-\delta g_{\rm e}$. In this approximation, the most unstable mode occurs at 
$k=0$. Qualitatively, an increase in $u$ will trigger an instability of the 
$k=0$ mode at a critical value $u_c$ when the linear equation first admits 
a non-zero solution for $\delta g_{\rm e}$. At the critical point, the 
corresponding fluctuation in the electron density then acquires the profile 
of the (normalized) zero energy eigenstate $\psi(x)$. Numerically one finds 
that the critical point for the instability occurs when $u_c/
\bar{g}_{\rm x}(0)=(2\ell/\ell_{\rm x})u_c \equiv a_0\simeq 6.516$, 
while the corresponding 
solution $\psi(x)$ is shown in Fig.~\ref{fig:wav_pot}.

While the approximation above identifies an instability, a perturbative 
analysis of the $k$-dependent corrections implied by (\ref{eq:xe}) reveals 
that the most unstable mode is spatially modulated. To the leading order of 
perturbation theory, an estimate of the shift of $u_c$ obtains 
\begin{eqnarray*}
\frac{\delta u_c}{u_c}=\frac{\ell}{a_0 a_1 }\left(
(k\ell)^2+\frac{a_0 a_2(Q)}{\ell_{\rm x}^2Q}\right)
\end{eqnarray*}
where $a_1=\int_{-\infty}^\infty dx\, \bar{g}_{\rm e}(x) \bar{g}_{\rm h}(x) 
\psi^2(x)\simeq 0.254\ell$, and 
\begin{eqnarray*}
a_2(Q)=\frac{1}{2}\int_{-\infty}^\infty dx dx'\, \bar{g}_{\rm e}(x) 
\bar{g}_{\rm h}(x) \psi(x) e^{-Q|x-x'|/\ell_{\rm x}} \psi(x').
\end{eqnarray*}
With $\ell_{\rm x}\gg \ell$, it will follow that $Q\ell \ll \ell_{\rm x}$, 
and the latter takes the constant value $a_2\simeq 0.461\ell^2$ independent 
of $k$.
%
%
Finally, minimizing $\delta u_c$ with respect to $k$, one finds that the 
instability occurs with a wavevector
\begin{equation}
k_c\ell_{\rm x}
\simeq \left(a_0 a_2\ell_{\rm x}/2\ell^3\right)^{1/3}
\label{eq:k_c}
\end{equation}
implying a shift of $u_c$ by 
%
%
$\delta u_c/u_c\sim (\ell/\ell_{\rm x})^{4/3}$. As a result, one can infer 
that the spatial modulation 
wavelength $\lambda_c\sim \ell^{1/3}\ell_{\rm x}^{2/3}$
is typically larger than the electron-hole overlap $\ell$, but smaller than  
$\ell_{\rm x}$.
Finally, 
an expansion of the nonlinear equations to higher order in fluctuations 
shows that, below the transition (i.e. for $u>u_c$), the amplitude of the 
Fourier harmonic $k_c$ grows as $(u-u_c)^{1/2}$. 

\begin{figure}[hbt]
\psfrag{amplitudeofmodulation}{Amplitude of Modulation}
\psfrag{u/gbar}{$u/\bar{g}_{\rm x}(0)$}
\psfrag{x/l}{$x/\ell$}
\psfrag{y/l}{$y/\ell$}
\psfrag{g,gx}{$g_{\rm x},\, g_{\rm e}$}
\begin{center}
\includegraphics[width=0.9\linewidth,angle=0]{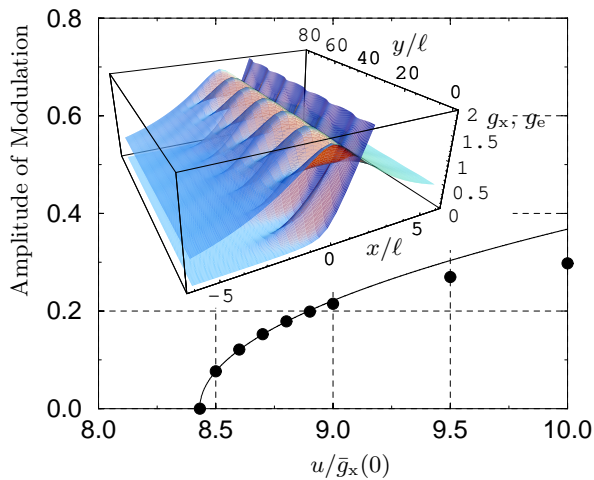}
\end{center}
\vspace{-0.2in}
\caption{Amplitude of the spatial modulation of the exciton density $g_x$ 
along the electron-hole interface $x=0$ as a function of the control 
parameter $u/\bar{g}_{\rm x}(0)$ as determined from the numerical solution 
of the nonlinear transport equation~(\protect\ref{eq:nontr}). 
A fit of the data to a square root dependence $\sim (u-u_c)^{1/2}$ is shown.
Inset: Dimensionless electron $g_{\rm e}$ and exciton $g_{\rm x}$ densities 
for $u/\bar{g}_{\rm x}(0)=9.5$ and $\ell_{\rm x}/\ell =4$. 
Here periodic boundary conditions are imposed along the interface.
} 
\label{fig:numerics}
\end{figure}

Once the instability is strongly developed (or when $\ell_{\rm x}\lesssim 
\ell$) the linear stability analysis above becomes unreliable. Here one must 
turn to numerics. Having in mind the mechanism of stimulated scattering, 
Fig.~\ref{fig:numerics} shows the results of a numerical analysis of the 
dimensionless nonlinear steady-state equation 
\begin{eqnarray}
\nabla_{\tilde {\bf x}}^2\ g_{\rm e}=
\exp\left(\frac{u}{\bar{g}_{\rm x}(0)}\delta g_{\rm x}\right)
g_{\rm e}(g_{\rm e}-\tilde x),
\label{eq:nontr}
\end{eqnarray}
where, using Eq.~(\ref{eq:xe}), $\delta g_{\rm x}\equiv g_{\rm x}-
\bar{g}_{\rm x}$ depends non-locally on $\delta g_{\rm e}\equiv g_{\rm e}-
\bar{g}_{\rm e}$ through the linear relation
\begin{eqnarray*}
\delta g_{\rm x}({\bf x})= 
-\delta g_{\rm e}({\bf x})+\int \frac{d^2{\bf x}'}{2\pi\ell_{\rm x}^2} 
K_0\left(\frac{|{\bf x}-{\bf x}'|}{\ell_{\rm x}}\right)
\delta g_{\rm e}({\bf x}'),
\end{eqnarray*}
and $K_0$ denotes the modified Bessel function. Although (as in the 
experimental ring geometry) the modulation is constrained by the periodic 
boundary conditions imposed along the direction parallel to the interface, 
the critical wavenumber $k_c$ lies close to that predicted by 
Eq.~(\ref{eq:k_c}). Similarly, the constraint leads to a value of 
$u_c/\bar{g}_{\rm x}(0)$ a little in excess of that predicted by the linear 
stability analysis. Finally, the amplitude of the instability confirms the 
square root dependence on $(u-u_c)$ predicted by perturbation theory.

In summary, we have shown that the realization of quantum degeneracy in a 
cold electron/hole-exciton system is signalled by the development of a 
spatial density modulation. Although our discussion is motivated by the 
photoexcited CQW system in which electrons and holes are spatially separated,
the mechanism is quite generic applying also to 
geometries where the electron and hole sources $J_{\rm e,h}$ have a
spatially independent profile.
By contrast, the instability mechanism appears to depend 
sensitively on there being a \emph{strongly} nonlinear dependence of the 
electron-hole binding rate on the exciton density pointing to the 
importance of stimulated scattering.

{\sc Acknowledgement:} We are indebted to Peter Littlewood, Alex Ivanov
and Daniel Chemla
for valuable discussions.



\end{document}